\documentclass[a4,11pt]{reportN}

\usepackage[latin1]{inputenc}   

\usepackage{graphics}
\usepackage{subfigure}
\usepackage{graphicx}
\usepackage{caption}
\usepackage{epsfig}
\usepackage[centertags]{amsmath}
\usepackage{graphicx,indentfirst,amsmath,amsfonts,amssymb,amsthm,newlfont}
\usepackage{longtable}
\usepackage{cite}
\usepackage[usenames,dvipsnames,svgnames,table]{xcolor}
\usepackage{float}
\usepackage[numbers]{natbib}
\usepackage{graphics}
\usepackage{subfigure}
\usepackage{graphicx}
\usepackage{multirow}
\usepackage{epsfig}
\usepackage[centertags]{amsmath}
\usepackage{graphicx,indentfirst,amsmath,amsfonts,amssymb,amsthm,newlfont}

\theoremstyle{remark}

\usepackage{longtable}
\usepackage{cite}
\usepackage{epstopdf}
\usepackage[hidelinks]{hyperref}
\usepackage[usenames,dvipsnames,svgnames,table]{xcolor}
\usepackage{verbatim}
\usepackage{listings}
\usepackage{algpseudocode,algorithm}

\usepackage{icomma}

\begin{document}

\title{Chaos suppression of Lorenz Systems by means on the average of rounding modes}

\author
    { \Large{Melanie Rodrigues e Silva}\thanks{me\_rodrigues\_silva@hotmail.com} \\
 {\small Group of Control and Modeling, UFSJ, S\~{a}o Jo\~{a}o del-Rei, MG, Brazil}\\
  \Large{Erivelton Geraldo Nepomuceno}\thanks{nepomuceno@ufsj.edu.br} \\
     {\small Department of Electrical Engineering, UFSJ, S\~{a}o Jo\~{a}o del-Rei, MG, Brazil}\\
   \Large{Samir Angelo Milani Martins}\thanks {martins@ufsj.edu.br} \\
 {\small Department of Electrical Engineering, UFSJ, S\~{a}o Jo\~{a}o del-Rei, MG, Brazil}}

\criartitulo


\begin{abstract}
{\bf Abstract}. This work deals with chaos suppression based on average of the rounded modes to negative and positive infinite. The present procedure acts to reduce the rounding errors. It was observed that when the method proposed in this paper is applied to the chaotic Lorenz's system, it exhibits a periodic behavior, characterized by a limit cycle and negative largest Lyapunov exponent. We tested our approach using three discretization schemes based on Runge-Kutta method.


\end{abstract}

\section{Introduction}

Studies and research related to dynamic systems are usually conducted by  computational analysis, which implies finite precision and the presence of error \cite{cobem}. According to the authors of  \cite{constantinides}, the use of computational arithmetic for the occurrence of scientific computation and engineering is indispensable and many iterative numerical algorithms can be considered dynamic systems. Generally, the algorithms are implemented in  a finite precision environment,  based on IEEE 754-2008 standard of the floating point \cite {IEEE}, which implies  generation of  error along the computational processes \cite{Nep2014}.

Among the many dynamic systems studied in literature, the Lorenz model is certainly one of the most well known. This model describes a meteorological system and it has been investigated mostly by its chaotic behaviour. The connection between finite precision and chaos may be done by the possibility of chaos control and chaos suppression. The former describes a strategy to eliminate chaos by means of an attachment of an auxiliary system which acts upon the system in order to stabilize in a desired state, for instance, a fixed point or a cycle limit \cite{OGY1990}, while the latter may be reached by means of changes on the discretization schemes or computer implementation without an external interference \cite{lorenz89}. There are many works on these issues, such as  \cite {jackson1991controls}, \cite{chen1994}, \cite {nijmeijer1995lyapunov}, \cite{chen1997control}, \cite{leung1997design}, \cite{chaossup}, \cite{Rajagopal2017}, \cite{peres}.

In this paper we present a new method to suppress chaos based on the average of round modes to negative and positive infinite, which can be seen as a performance of filter as proposed by \cite{chua}. The proposed method acts in reducing  error due to the computational rounding processes and to the finite precision. We use as paradigm the Lorenz's system and tested our approach using three discretization schemes based on Runge-Kutta method of $3^{th}$, $4^{th}$ and $5^{th}$ order.
\medskip

\section{Methodology}

Consider the following definitions.\\

\noindent \textbf{Definition 1.} \emph{An orbit is a sequence of values of a nonlinear dynamic system, represented by ${x_n} = [x_0,x_1,...,x_n]$} \citep{nep2016}.\\

\noindent \textbf{Definition 2.} \emph{Let $i \in \mathbb{N}$ represents a pseudo-orbit, which is defined by an initial condition and a combination of software and hardware. A pseudo-orbit is an approximation of an orbit and can be represented as}
	\begin{equation}
	\{\hat{x}_{i,n}\}=[\hat{x}_{i,0},\hat{x}_{i,1},\dots,\hat{x}_{i,n}]   
	\end{equation}
	such that,
	\begin{equation}
	\mid x_n - \hat{x}_{i,n} \mid \le \xi_{i,n}
	\end{equation}
	\emph{where $\xi_{i,n} \in \mathbb{R}$ is the limit of error and $\xi_{i,n} \ge 0$} \citep{nep2016}.\\


According to \textit{Definition 2.}, pseudo-orbits are approximations of true orbits, so such approximations incorporate uncertainties into the simulations of nonlinear dynamic systems. Therefore, it is known that errors related to rounding processes and hardware limitations are recurrent in simulations, in order to reduce $\xi_{i,n}$ Silva et al., \cite {chua}, proposed a filter type based on the average of the rounded modes. \\ 

\noindent \textbf{Preposition 1.} \emph{
	Let $\hat{{x}}_{i,n}^-$ and ${\hat{x}}_{i,n}^+$ be the calculated pseudo-orbit value by round towards positive infinite and round towards negative infinite, respectively. The arithmetic average given by 
	\begin{equation}
	\hat{x}_{j,n}=\frac{\hat{{x}}_{i,n}^+ + {\hat{x}}_{i,n}^-}{2}
	\label{average}
	\end{equation}
	is an alternative pseudo-orbit.}\\    

Let Lorenz model be  described by a set of three differential equations as show in Equation \ref{lorenz}, where $ \sigma $ and $ r $ are called Prandtl and Rayleigh numbers, respectively. The state $ x $ represents the intensity of the convection movement, $ y $ is proportional to the temperature variation between upstream and downstream currents and $ z $ is proportional to the distortion of the vertical temperature profile \citep{lorenz1963deterministic}.

\begin{equation}
\left\{ \begin{array}{rcl}
\dot{x}  &=& \sigma (y-x) \\
\dot{y} &=&  x (\rho - z) - y \\
\dot{z} &=& xy -\beta z  
\end{array}\right.
\label{lorenz}
\end{equation}

\begin{table}[!ht]
	\centering \caption{\label{tab:1}Values of initial conditions and the constants of time for Lorenz's system.}
	\begin{tabular}{|c|c|}				\hline 	
		\textbf{Parameters} &  \textbf{Values} \\ \hline 
		$x_0$ &   0.06735 \\ \hline 
		$y_0$ &   1.8841 \\ \hline
		$z_0$ &   15.7734 \\ \hline
		Integration step & 10 $ms$    \\ \hline
		Time of simulation & 100 $s$    \\ \hline
	\end{tabular}
\end{table}

\begin{table}[!ht]
	\centering \caption{\label{tab:system} Features of Lorenz's system.}
	\begin{tabular}{|c|c|}				\hline 	
		\textbf{Parameters} &  \textbf{Values} \\ \hline 
		$\sigma$ & 7.6    \\ \hline
		$r$ & 65    \\ \hline
		$b$ & 5.3    \\ \hline
	\end{tabular}
\end{table}
The algorithms set forth in the \textit {Appendice} for the Lorenz's system were elaborated in \textit{\textbf {Matlab}} software. The initial condition, integration step and simulation time are  shown in Table \ref{tab:1}, and other parameters in Table \ref{tab:system}. Algorithm 1 represents the solution of differential equations under Runge-Kutta method of 3, 4 and 5 order. In Algorithm 2, the average of the rounded modes is performed and the \textit{Prepostion 1.} is applied by means of the Matlab function $ system\_dependent $.


According to the conditions of Table \ref{tab:1} and Table \ref{tab:system}, it is expected that the intensity of convection movement to be chaotic, such behavior is verified in Figure 1a, Figure 1c and Figure 1e. This solution is carried out by the  classical system simulation method of Runge-Kutta of fourth order, according to IEEE 754-2008 standard.

Using the proposed method, we obtain the solutions for each variable presented in  Figure 1b, Figure 1d and Figure 1f. The difference is clear and now the system performs a periodic behaviour. The calculation of the largest Lyapunov exponent Kodba et al. \cite{kodba} by \citep{Lyapunov} has been performed and reinforces this evidence, showing that the Lorenz model solved by the proposed method presents a periodical behaviour, as the Lyapunov exponent is negative, as shown in Table \ref{tab:parametros}.

Therefore, the results of the classical simulations for the three species of discretization, under IEEE 754-2008, are the result of a chaotic behavior. But, the solution using the proposed method is  periodic.

\begin{figure}[!ht]
\begin{center}
		\includegraphics[width=1\linewidth]{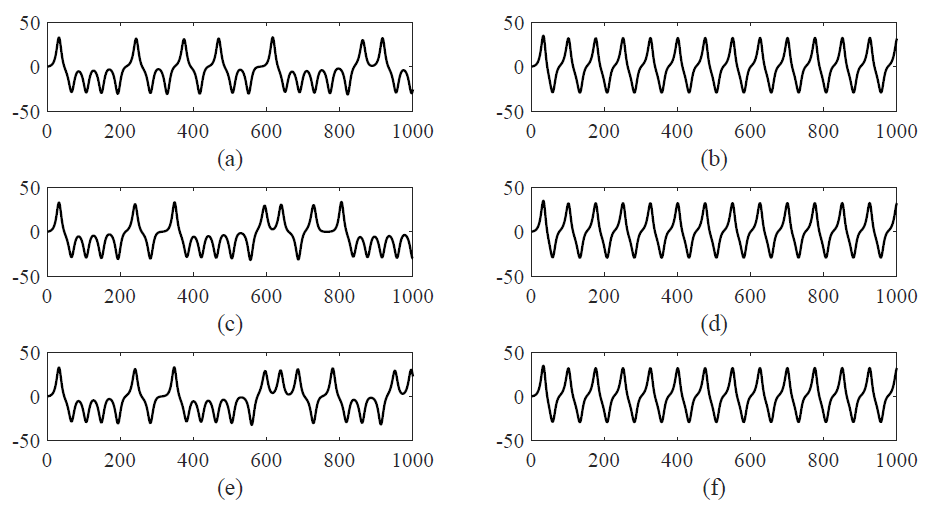}
		\caption{Intensity of the convection motion by iterations (a) Traditional simulation realised by RK3 method. (b) Suppressed chaos method proposed in this paper under RK3 method. (c) Traditional simulation realised by RK4 method. (d) Suppressed chaos method proposed in this paper under RK4 method. (e) Traditional simulation realised by RK5 method. (e) Suppressed chaos method proposed in this paper under RK5 method.} 
			\label{fig:waves}
		\end{center}
	\end{figure}

\begin{table}[!ht]
		\centering \caption{\label{tab:parametros}The largest Lyapunov exponent of Lorenz system for both methodology.}
		\begin{tabular}{|c|c|c|}
			\hline 	
			Discretization Method & \textbf{Traditional Method} &  \textbf {Proposed Method}
			\\ \hline RK3 &  0.344215 &  -0.011370
			\\ \hline RK4 &  0.087915 &  -0.001371
			\\ \hline RK5 &  0.165580 &  -0.001362\\ \hline
		\end{tabular}
	\end{table}
	
\section{Conclusion}

This paper presented a new method to show chaos suppression based on the average of rounding modes. We used three different discretization schemes to solve the Lorenz model. Using an approach based on Runge-Kutta of third, fourth and fifth order, the simulations exhibits the expected chaotic behaviour. On the other hand, if we take the average of two rounding modes for each step, the model presents a periodical behaviour, proved by their negative Lyapunov exponent. This phenomenon has been already reported to Chua's circuit in a similar approach \cite{SNAM2017}. It is expected that the average could reduce the variance of error. If so, why a lower variance noise result would present a periodical behaviour? We believe that this deservers further investigation. As said \cite{Resenha}, small neglected physical effects on mathematical models, or even simple rounding errors, can be amplified over time so that the predicted motion differs from real motion.

\subsection*{Acknowledgements}

This work has been supported by the CNPq and CAPES. A special thanks to Control and Modelling Group (GCOM).

\section{Appendix}
	
\subsection{Algorithm 1}	
\vspace{-0.5cm}
\begin{lstlisting} 
system_dependent('setround',-Inf);
ym=[0.06735 1.8841 15.7734];
system_dependent('setround',Inf);
yp=[0.06735 1.8841 15.7734];
system_dependent('setround',0.5);
y1=(ym+yp)/2;
ym=y1;
yp=y1;
tf=100; 
h=1e-2;
tspan = 0:h:tf;
N=length(tspan);
for k=1:N-1
system_dependent('setround',-Inf);
aux = ode(@lorenz,tspan(k:k+1),ym(k,:),yp(k,:));
ym(k+1,:)=aux(2,:);
system_dependent('setround',Inf);
aux = ode(@lorenz,tspan(k:k+1),yp(k,:),ym(k,:));
yp(k+1,:)=aux(2,:);
end		
figure()
plot(1:1000,ym(1:1000,1),'k',1:1000,
yp(1:1000,1),'k', 'Linewidth',5)
label('Iteracoes')
ylabel('x Lorenz')
set(gcf, 'PaperPosition', [0 0 20 10]); 
set(gcf, 'PaperSize', [50 10]); 
set(gca,'FontSize',30,'Fontname','Times New Roman')
title('Lorenz');
saveas(gcf, 'Lorenz', 'pdf')
%better perspective
view(-16,24)
grid
saveas(gcf,'Lorenz_atrator','pdf')
\end{lstlisting}
		
\subsection{Algorithm 2}
\vspace{-0.5cm}
\begin{lstlisting} 			
function out = lorenz(t,in,in2)
%Recive the inputs:
x = in(1); 
y = in(2); 
z = in(3); 
x2=in2(1);
y2=in2(2);
z2=in2(3);
% The average of two inputs:
system_dependent('setround',0.5);
x=(x+x2)/2;
system_dependent('setround',0.5);
y=(y+y2)/2;
system_dependent('setround',0.5);
z=(z+z2)/2;
% System parameters
sigma = 7.6;r = 65;  b = 5.3; 
% Diferential equations
xdot = sigma*(y - x);
ydot = x*(r - z) - y;
zdot = x*y - b*z;
out = [xdot ydot zdot]';
\end{lstlisting}

\end{document}